\documentclass[english,floatfix,superscriptaddress,nofootinbib,prd]{revtex4}
\pdfoutput=1
\usepackage[paperwidth=210mm,paperheight=297mm,centering,hmargin=2cm,vmargin=2cm]{geometry}
\usepackage[utf8]{inputenc}
\usepackage[T1]{fontenc}
\usepackage{lmodern}
\usepackage{amsmath}
\usepackage{amssymb}
\usepackage{graphicx,epsfig}
\usepackage{esint}
\usepackage{dcolumn}
\usepackage{xcolor} 
\usepackage{csquotes}
\usepackage{marvosym}
\usepackage{hyperref}
\usepackage[overload]{textcase}
\usepackage[symbol,hang,flushmargin]{footmisc}

\DefineFNsymbolsTM{otherfnsymbols}{%
  \Radioactivity	\circ
  \Yinyang   \mathsection
  \Biohazard    \dagger
  \Bat \ddagger
}%

\setfnsymbol{otherfnsymbols}

\newenvironment{seqn}{\equation\aligned}{\endaligned\endequation}

\newcommand{\be}{\begin{seqn}}
\newcommand{\ee}{\end{seqn}}

\newcommand{\bea}{\begin{eqnarray}}
\newcommand{\eea}{\end{eqnarray}}

\newenvironment{arabicfootnotes}
  {\par\edef\savedfootnotenumber{\number\value{footnote}}
   
   \setcounter{footnote}{0}}
  {\par\setcounter{footnote}{\savedfootnotenumber}}

\begin{document}
%
%
%
%
%
%
\title{Linear and Quadratic GUP, Liouville Theorem, Cosmological Constant, and Brick Wall Entropy}

\author{Elias~C.~Vagenas}
\email{elias.vagenas@ku.edu.kw}
\affiliation{Theoretical Physics Group, Department of Physics, Kuwait University, P.O. Box 5969, Safat 13060, Kuwait.}

\author{Ahmed~Farag~Ali}
\email{ahmed.ali@fsc.bu.edu.eg; ahmed@quantumgravityresearch.org}
\affiliation{Department of Physics, Faculty of Science, Benha University, Benha, 13518, Egypt.}
\affiliation{Quantum Gravity Research, Los Angeles, CA 90290, USA.}

\author{Mohammed~Hemeda}
\email{mhemeda@sci.asu.edu.eg}
\affiliation{Department of Mathematics, Faculty of Science, Ain Shams University,  Cairo, 11566, Egypt.}

\author{Hassan~Alshal}
\email{halshal@sci.cu.edu.eg}
\affiliation{Department of Physics, Faculty of Science, Cairo University, Giza, 12613, Egypt.}
\affiliation{Department of Physics, University of Miami, Coral Gables, FL 33146, USA.}
\begin{abstract}
\par\noindent
Motivated by the works  on Equivalence Principle in the context of  linear Generalized Uncertainty Principle  and, independently, in the context of  quadratic Generalized Uncertainty Principle,  we expand these endeavors in the context of Generalized Uncertainty Principle when both linear and quadratic terms in momentum are include. We demonstrate how the definitions of equations of motion change upon that expansion. We also show how to obtain an analogue of Liouville theorem in the presence of linear and quadratic Generalized Uncertainty Principle. We employ the corresponding modified invariant unit volume of phase space to discuss the resulting density of states, the problem of cosmological constant, the black body radiation in curved spacetime, the concurrent energy and consequent no Brick Wall entropy. 
\end{abstract}
\maketitle
\begin{arabicfootnotes}
%
%
%
%
\section{Introduction}
%
%
%
%
\par\noindent
As a consequence of perturbative string theory, modifying the standard Heisenberg Uncertainty Principle (HUP) into the Generalized Uncertainty Principle (GUP), by adding an extra quadratic term in momentum, resulted in proposing that gravity might behave differently at the minimal length scale  compared with how it does in general relativity \cite{Veneziano:1986zf, Gross:1987ar, Amati:1988tn, Konishi:1989wk, Maggiore:1993rv, Garay:1994en, Scardigli:1999jh, Adler:2001vs}. After this proposal, in a series of papers  \cite{Das:2008kaa, Das:2009hs, Ali:2009zq, Ali:2010yn, Das:2010sj, Das:2010zf, Ali:2011fa} the first two authors of this paper, namely ECV and AFA, together with Saurya Das, introduced a linear and quadratic GUP (LQGUP), i.e., GUP with linear and quadratic terms in momentum, in such a way that uncertainty principle becomes compatible with Doubly Special Relativity (DSR) theories \cite{Colella:1975dq, Magueijo:2001cr, Magueijo:2004vv, Cortes:2004qn} and consistent with commutation relations of phase space coordinates $[x_i,x_j]=[p_i,p_j]=0$ via Jacobi identity. In Ref. \cite{Ali:2009zq}, the commutation relation becomes
\be\label{GUPcomm}
\left[x_i, p_j\right]=i\hbar\left[\delta_{ij}-\alpha\left(\delta_{ij}~p+\frac{p_i p_j}{p}\right)+\alpha^2\left(\delta_{ij}~p^2 +3p_i p_j\right)\right]~.
\ee
\par\noindent
In addition, this commutation relation is also associated to the outcome of  a  \emph{perturbative} solution, up to third order, $\psi\sim e^{ix/\Delta x_{min}}$ of Schr\"{o}dinger equation such that it is endowed with a periodic nature of minimal length $\Delta x_{min}=\alpha_0 \ell_p$, suggesting that spacetime has a discrete nature \cite{Ali:2009zq}. Earlier before that, Chang {\it et al.} \cite{Chang:2001bm} used the quadratic GUP (QGUP), i.e., GUP with a quadratic term in momentum, to study its effect on the UV/IR momentum behavior and the implications on density of states and the cosmological constant problem \footnote{The cosmological constant problem has also been discussed in the context of the LQGUP-deformed Wheeler-DeWitt equation \cite{Garattini:2015aca}.}. They concluded that holography in a cosmological background might introduce another scale other than $\frac{1}{\alpha_0\ell_p}$ due to the suppressed density of states in UV case. Therefore, the number of degrees of freedom contributing to the vacuum energy density would be very small.
Following this line of research, one of the authors, namely AFA, did the same calculations \cite{Ali:2011ap} upon considering only the linear GUP (LGUP), i.e.,  GUP with a linear term in momentum. 
The linear term in momentum of LGUP changes the power of the unit volume of phase space from $D$, as in Ref. \cite{Chang:2001bm} to $D+1$, but  it does not suppress the density of states. Therefore, the effect of LGUP on holographic entropy of the cutoff phase space disagrees with 't Hooft's standard result, that forces disagreement between the micro-canonical and canonical ensembles for such system with large number of degrees of freedom.
\par
The rest of this work is structured as follows. In  section \textbf{II} we reconsider the effect of LQGUP on the equivalence principle and the equations of motions . In section \textbf{III} we examine the effect of LQGUP on the unit volume of phase space, and whether we should consider the correction factor to be raised to power $D$ or $D+1$.Then, in section \textbf{IV} we see the consequences  on the cosmological constant problem. Moreover, in section \textbf{V} we investigate the outcome of introducing LQGUP to energy distribution of massless black body radiation. In addition, in section \textbf{VI}, we compare the effect of LQGUP with the effect of LGUP and QGUP on massless particles in general static spherically symmetric curved spacetime. Furthermore, in section \textbf{VII} we introduce LQGUP to the Brick Wall entropy of black holes. Finally, we discuss the contrasts and similarities among the different orders of GUPs and, therefore, conclude We take the units $G=c=\hbar=k_B=1$.
%
%
%
%
%
\section{LQGUP Equivalence Principle and Equations of Motion}
%
%
%
%
%
\par\noindent
For the classical limit of Eq.(\ref{GUPcomm}) of any two canonical conjugates $\hat{P}$ and $\hat{Q}$, the correspondence principle states that%
\be\label{corresp}
\frac{1}{i\hbar} [\hat{P},\hat{Q}] \rightarrow \{P,Q\}
\ee
\par\noindent
where the square brackets stand for the Lie brackets  while the curly ones stand for Poisson brackets. Meanwhile the relation between the expectation value of any QM observable and the expectation value  of the commutator of that observable with the Hamiltonian of the system is given by
\be\label{expect}
\frac{d}{dt}\Big\langle A \Big\rangle = \frac{1}{i\hbar}\bigg\langle \big[A,H\big] \bigg\rangle + \left\langle \frac{\partial}{\partial t}A \right\rangle~.
\ee
\par\noindent
Upon employing the correspondence principle, as stated in Eq.(\ref{corresp}), on Eq.(\ref{expect}) for position, we obtain
\be\label{xH}
\dot{x_i} = \{ x_i,H \} = \delta_{ij} \frac{\partial H}{\partial p_j} = \{ x_i,p_j \} \frac{\partial H}{\partial p_j}
\ee
and for the momentum we get
\be\label{pH}
\dot{p_j}=-\{x_i,p_j\} \frac{\partial V}{\partial x_j}~.
\ee
\par\noindent
Then, we utilize Eq.(\ref{GUPcomm}) in the above two expressions to get
\be
& \dot{x} =(1-2\alpha p +4\alpha^{2} p^2)\frac{p}{m}\\
& \dot{p} = -(1-2\alpha p +4\alpha^{2} p^2)\frac{\partial V}{\partial x}~.
\ee
\par\noindent
Consequently, the definition of the  force reads
\be
F= m\ddot{x} & = m\{\dot{x},H\}\\
& = (1-4\alpha p + {12} \alpha^2 p^2)\{p,H\}\\
& = -(1- 4\alpha p+{12}\alpha^2 p^2)(1-2\alpha p + 4\alpha^2 p^2) \frac{\partial V}{\partial x}\\
& = - \left[1-6\alpha p + {24}\alpha^2 p^2 + \mathcal{O}(\alpha^3)\right]\frac{\partial V}{\partial x}~.
\ee
\par\noindent
It is noteworthy that $p$ and $F$ are no longer equal to $m\dot{x}$ and $-\partial V/\partial x$, respectively. The $\alpha$ term matches with the results obtained in Ref. \cite{Ali:2011ap}. In addition, we have an  $\alpha^2$ term,  as expected, and  this   $\alpha^2$ term does not contradict the conclusion about the dynamical violation of equivalence principle obtained in Ref. \cite{Ali:2011ap}. LQGUP controls the UV divergences such that it shows similar cosmological implications of the dark sector where the associated long-range force acts only between nonbaryonic particles \cite{Keselman:2009nx}. It should be stressed that the violation of equivalence principle  obtained here also agrees  with that obtained from tidal forces in the domains of string theory \cite{Mende:1992pm, Garay:1994en}.
%
%
%
%
%
\section{LQGUP and Liouville Theorem}
%
%
%
%
%
\par\noindent
In the light of Eq.(\ref{GUPcomm}), it is evident  the momentum dependence of the unit volume of each quantum state in the phase space. This would contradict that laws of physics should not change their form with respect to any  change in space and time, i.e., the unit volume of the space has to be invariant upon the change in the momentum for every state.  Therefore, we look for an analogue to Liouville theorem by assuming the change in position and momentum in a time $\delta t$ as
\be\label{x'p'}
& x'_i = x_i + \delta x_i = x_i + \dot{x_i}\delta t +\mathcal{O}(\delta t^2)\\
& p'_i = p_i + \delta p_i = p_i + \dot{p_i}\delta t +\mathcal{O}(\delta t^2)~.
\ee
We demand the Jacobian --which relates the states of phase space before and after a time $\delta t$-- to be 
\be \label{Jacobian}
\bigg\vert \frac{ \partial ( x'_1 ,\cdots , x'_D ; p'_1, \cdots, p'_D)}{\partial (x_1 ,\cdots , x_D ; p_1, \cdots, p_D )} \bigg\vert = 1+\left(\frac{\partial \delta x_i}{\partial x_i}+\frac{\partial \delta p_i}{\partial p_i}\right)+\cdots
\ee
such that the phase space volume element after $\delta t$ becomes
\be \label{phase}
d^D \textbf{x}' d^D \textbf{p}'= \bigg\vert \frac{ \partial ( x'_1 ,\cdots , x'_D ; p'_1, \cdots, p'_D)}{\partial (x_1 ,\cdots , x_D ; p_1, \cdots, p_D )} \bigg\vert d^D\textbf{x} d^D\textbf{p}~.
\ee
Upon combining Eqs.(\ref{GUPcomm}), (\ref{xH}), (\ref{pH}),  and (\ref{x'p'}), we express the variation term in the RHS of Eq.(\ref{Jacobian}) as
\be \label{varJacob}
\bigg( \frac{\partial \delta x_i}{\partial x_i} +\frac{\partial \delta p_i}{\partial p_i}\bigg) = -\frac{\partial}{\partial p_i} \bigg[ \delta_{ij}-\alpha\left(\delta_{ij}~p+\frac{p_i p_j}{p}\right) + \alpha^2\left(\delta_{ij}~p^2 +3p_i p_j\right)\bigg] \frac{\partial H}{\partial x_j}\delta t
\ee
where again the $\alpha$ term matches with the one in Ref. \cite{Ali:2011ap} and is evaluated there to be
\be \label{LinAlpha}
-\frac{\partial}{\partial p_i} \left[-\alpha\left(\delta_{ij}~p+\frac{p_i p_j}{p}\right) \right] = \alpha (D+1) \frac{p_i}{p}~,
\ee
meanwhile the $\alpha^2$ term is evaluated as
\be \label{QuadAlpha}
-\frac{\partial}{\partial p_i} \left[ \alpha^2\delta_{ij}~p^2 +3p_i p_j\right] = -2\alpha^2 (D+1) \left[1+ \frac{2}{D+1}\right] p_i~.
\ee
Now we substitute Eqs.(\ref{LinAlpha}) and (\ref{QuadAlpha}) in the RHS of Eq.(\ref{Jacobian}) to get
\be\label{expandJacob}
1+\left(\frac{\partial \delta x_i}{\partial x_i}+\frac{\partial \delta p_i}{\partial p_i}\right) = 1+(D+1)\left[\frac{\alpha}{p}-2\alpha^2-\frac{4\alpha^2}{D+1}\right]p_j\frac{\partial H}{\partial x_j}\delta t~.
\ee
To obtain the correct scale factor that makes LQGUP compatible with Liouville theorem, we consider the infinitesimal time evolution in the linear term to the first order in $\alpha$ and $\delta t$ from  Ref. \cite{Ali:2011ap} as
\be\label{LinP'}
(1-\alpha p') \sim (1-\alpha p) \left[1+\alpha\frac{p_i}{p}\frac{\partial H}{\partial x_j}\delta t\right]~,
\ee
and the infinitesimal time evolution in the quadratic term to the second order in $\alpha$ 
and first order in $\delta t$ as
\be \label{QuadP'}
\alpha^2(\frac{2}{D+1}+\frac{1}{2})p'^2 & \sim \alpha^2(\frac{2}{D+1}+\frac{1}{2})\left(p^2+2p_i\delta p_i\right)\\
& \sim \alpha^2(\frac{2}{D+1}+\frac{1}{2})\left[p^2-2p_i\{ x_i,p_j \} \frac{\partial H}{\partial x_j}\delta t\right]\\
& \sim \alpha^2(\frac{2}{D+1}+\frac{1}{2})\left(p^2-2p_i\delta_{ij} \frac{\partial H}{\partial x_j}\delta t\right)+\mathcal{O}(\alpha^3)\\
& \sim \alpha^2(\frac{2}{D+1}+\frac{1}{2})\left(p^2-2p_j \frac{\partial H}{\partial x_j}\delta t\right)~.
\ee
\par\noindent
Then, we combine Eq.(\ref{LinP'}) and Eq.(\ref{QuadP'}) to get
\be \label{LinQuadP'}
1-\alpha p' + \alpha^2\left(\frac{2}{D+1}+\frac{1}{2}\right)p'^2 \sim 1-\alpha p+ \alpha^2\left(\frac{2}{D+1}+\frac{1}{2}\right)p^2 + \left[\frac{\alpha}{p}(1-2\alpha p)-\alpha^2-\frac{4\alpha^2}{D+1}\right]p_j\frac{\partial H}{\partial x_j}\delta t~.
\ee
\par\noindent
We factor out $\left[1-\alpha p+ \alpha^2\left(\frac{2}{D+1}+\frac{1}{2}\right)p^2\right]$ in the RHS such that Eq.(\ref{LinQuadP'}) becomes

\bea
1-\alpha p' + \alpha^2\left(\frac{2}{D+1}+\frac{1}{2}\right)p'^{2}\!\!\!&\sim&\!\!\!  \left[1-\alpha p+ \alpha^2\left(\frac{2}{D+1}+\frac{1}{2}\right)p^2\right]\nonumber\\
\!\!\!&&\!\!\!\times \Bigg[ 1+\frac{(1-2\alpha p)~(\alpha/p)}{\left(1-\alpha p+ \alpha^2\left(\frac{2}{D+1}+\frac{1}{2}\right)p^2\right)} -\frac{\alpha^2+ (4\alpha^2 / (D+1))}{\left(1-\alpha p+ \alpha^2\left(\frac{2}{D+1}+\frac{1}{2}\right)p^2\right)}\Bigg]p_j \frac{\partial H}{\partial x_j}\delta t\nonumber\\
\!\!\!& \sim&\!\!\! \left[1-\alpha p+ \alpha^2\left(\frac{2}{D+1}+\frac{1}{2}\right)p^2\right] \nonumber\\
\!\!\!&&\!\!\! \times \Bigg\{1+\left[\frac{\alpha}{p}(1-2\alpha p)(1+\alpha p)-\alpha^2 - 4\frac{\alpha^2}{D+1} + \mathcal{O}(\alpha^3)\right]p_j\frac{\partial H}{\partial x_j}\delta t\Bigg\}~.
\eea
Or,
\bea
1-\alpha p' + \alpha^2\left(\frac{2}{D+1}+\frac{1}{2}\right)p'^{2} \!\!&\sim&\!\! \left[1-\alpha p + \alpha^2\left(\frac{2}{D+1}+\frac{1}{2}\right)p^2\right] \nonumber\\
\!\!\!&&\!\!\!\times\left[ 1+\left(\frac{\alpha}{p}-2\alpha^2 - \frac{4\alpha^2}{D+1}+\mathcal{O}(\alpha^3)\right)p_j\frac{\partial H}{\partial x_j}\delta t \right]~.
\eea
Finally, we raise the last result to power $-(D+1)$ then expand it to the first order of binomial coefficient such that the \emph{weight factor} of LQGUP, that corrects the definition of unit volume phase space, is defined as
\bea \label{monster}
\left(1-\alpha p' + \alpha^2\left(\frac{2}{D+1}+\frac{1}{2}\right)p'^2\right)^{-(D+1)} &\sim& \left[1-\alpha p + \alpha^2\left(\frac{2}{D+1}+\frac{1}{2}\right)p^2\right]^{-(D+1)}\nonumber\\
&&\times \left[ 1-(D+1)\left(\frac{\alpha}{p}-2\alpha^2 - \frac{4\alpha^2}{D+1}\right)p_j\frac{\partial H}{\partial x_j}\delta t \right].
\eea
\par\noindent
By comparing Eq.(\ref{expandJacob}) with Eq.(\ref{monster}), the corrected LQGUP invariant-under-time unit volume of phase space  is given by
\be\label{volume}
\frac{d^D\mathbf{x} d^D\mathbf{p}}{(2\pi)^D \left[1-\alpha p + \left(\frac{2}{D+1}+\frac{1}{2}\right)\alpha^2 p^2\right]^{(D+1)}}
\ee
\par\noindent
which, technically, will later define the number of quantum states per momentum space volume upon integrating over $d^D\mathbf{x}$. Consequently, this would affect the calculations of energy, holographic entropy, and cosmological constant. Before we discuss these, we want to emphasize on the different results obtained  in Refs. \cite{Chang:2001bm, Ali:2011ap}. In Ref. \cite{Chang:2001bm}, the power that appears in the corresponding equation to Eq.(\ref{volume}) is not $(D+1)$ but $D$  and, in addition, there is no $\alpha$ term. In Ref. \cite{Ali:2011ap}, it has the same power as we have even even if it does not have the $\alpha^2$ term. Since $\alpha^2\sim\beta$, where $\beta$ is the minimal length factor in Ref. \cite{Chang:2001bm},  we expect the behavior of LQGUP weight factor to be close to that of Ref. \cite{Chang:2001bm}, as shown in Fig. 1. However, the computational results are quite different, due to the divergent behavior of the linear term we have, as  we will see in next sections. This numerical difference between QGUP and LQGUP is crucial when we consider the quantum gravity effects within the vicinity of the minimal length.
\begin{figure}[h!]
  \includegraphics[width=0.5\textwidth]{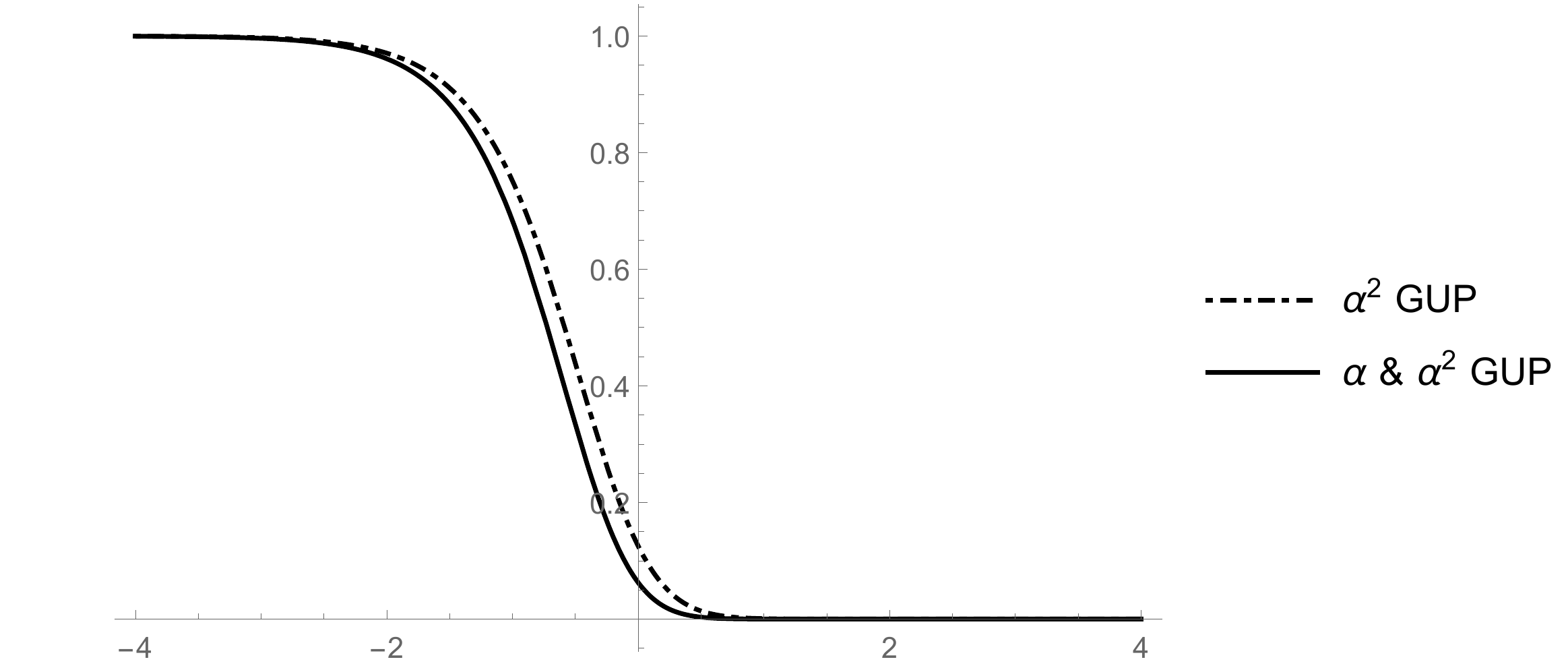}
  \caption{The behavior of weight factor $(1-\alpha p +\alpha^2 p^2)^{-4}$ of LQGUP compared to $(1+\beta p^2)^{-3}$ of Ref. \cite{Chang:2001bm} when D=3. The horizontal axis is the logarithm of every weight factor. We have set $\alpha^2=\beta=1$.}
  \label{fig.1}
\end{figure}
%
%
%
%
%
%
%
\section{LQGUP effect on Cosmological Constant}
%
%
%
%
%
\par\noindent
Based on the LQGUP analogue of Liouville theorem derived in the previous section, the sum over all harmonic oscillator momentum states per unit volume will now read
 \footnote{The numerical factor $2\pi$ in front of the integral of $\Lambda_{\text{LQGUP}}$ in Eq.(\ref{Lambda}) should have been $1/2\pi^2$ as it is in next section  (see Eq.(\ref{statesNo})). However, for the sake of comparison between our result given here and the result for QGUP obtained in Ref.  \cite{Chang:2001bm}, we keep it $2\pi$ .}
\be\label{Lambda}
\Lambda(m)= 2\pi\int \limits_0^{\infty}\frac{p^2}{(1-\alpha p + \alpha^2 p^2)^4}\sqrt{p^2+m^2}\, dp~.
\ee
Upon considering $\displaystyle{\tan\theta=\frac{2\alpha p-1}{\sqrt{3}}}$, the above integral becomes
\be
\Lambda(m)=2\pi \left(\frac{4}{3}\right)^4 \frac{\sqrt{3}}{2\alpha} \int \limits_{-\pi/6}^{\pi/2} \cos^6\theta \left(\frac{\sqrt{3}\tan\theta +1}{2\alpha}\right)^2\left[\left(\frac{\sqrt{3}\tan\theta +1}{2\alpha}\right)^2+m^2\right]^{1/2} d\theta~.
\ee
This integral is not easy to be exactly solved. However, we still can compare our result here with those obtained in Refs. \cite{Chang:2001bm,Ali:2011ap}. This is done in Fig. 2, after setting $m=0$ \footnote{In the context of Gravity's Rainbow, a similar plot was obtained in Ref. \cite{Garattini:2011kp}.}. The massless cosmological constant corresponding to LQGUP reads
\be \label{Lambda0}
\Lambda_{\text{LQGUP}}(0)= 2\pi \frac{\sqrt{3}}{2\alpha} \left(\frac{4}{3}\right)^4 \frac{27\sqrt{3}+28\pi}{384\alpha^3} \sim \frac{2\pi}{\alpha^4}
\ee
which is much larger than the $\Lambda_{\text{QGUP}}(0)$ obtained in Ref. \cite{Chang:2001bm} \footnote{Remember that, in Ref. \cite{Chang:2001bm}, $\beta \sim \alpha^2$.}.
It is easily seen that  the cosmological constant of LQGUP is still finite with $\alpha$ and $\alpha^2$ to be the UV cutoff. However, we agree with Chang et al. in Ref. \cite{Chang:2001bm} that this does not resolve the cosmological constant problem since $\alpha^{2}\sim M_P$ with $M_P $ to be the Planck mass.

\begin{figure}[h!]
  \includegraphics[width=0.5\textwidth]{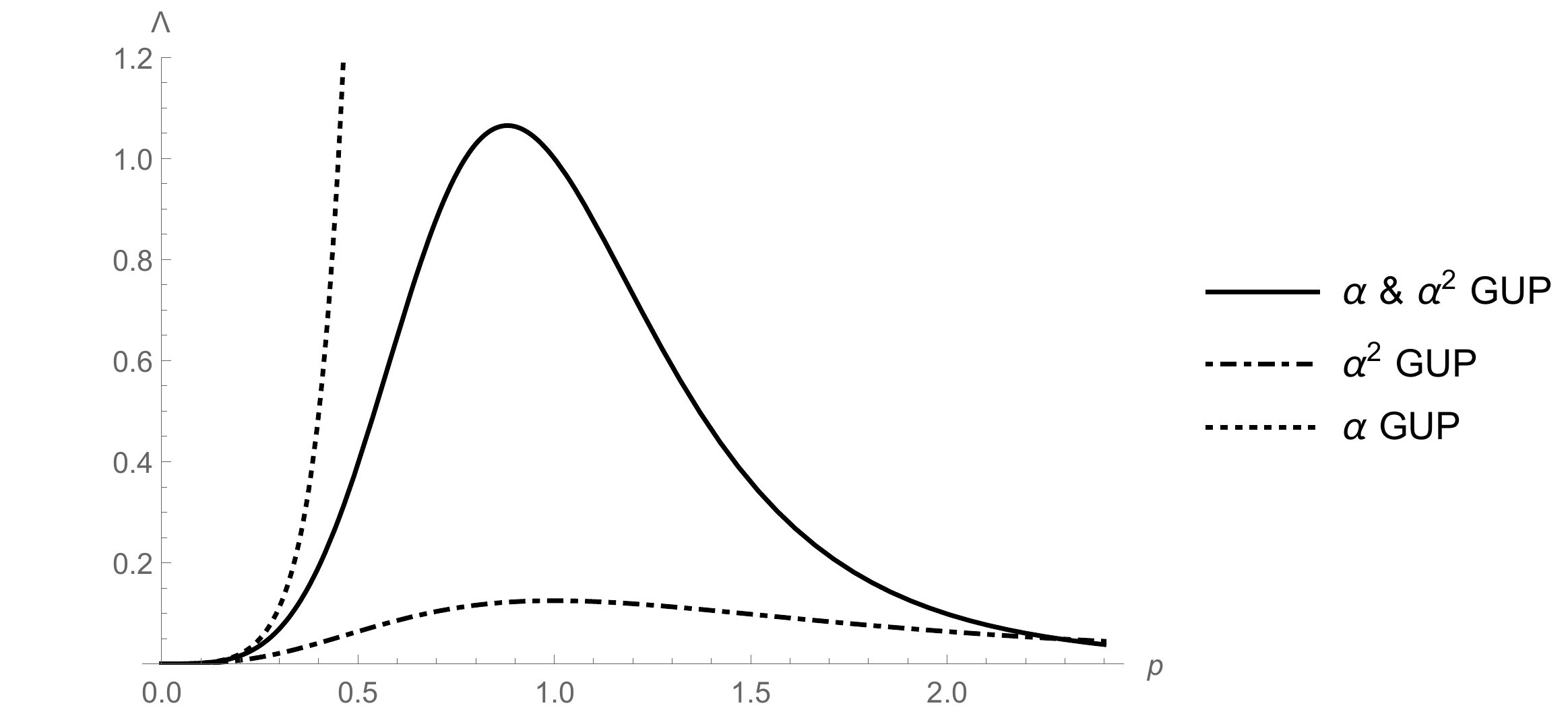}
  \caption{The effect of different weight factors on the calculations of cosmological constant upon considering LGUP, QGUP and LQGUP when D=3. We have set $\alpha=1$ and $m=0$.}
  \label{fig.2}
\end{figure}
%
%
%
%
%
\section{LQGUP effect on Energy Distribution of Black Body Massless Radiation}
%
%
%
%
%
\par\noindent
In this section, we calculate the energy distribution of blackbody massless radiation in the framework of LQGUP.
First, we set $m=0$ so that for the single massless particle we get $E=\sqrt{p^2+m^2}=p$. Then, the total number of quantized modes for massless bosonic field in a cubic box (with $D=3$) of size $L$ reads
\be \label{statesNo}
N = & \frac{L^3}{2\pi^2}\int\limits_0^{\infty} \frac{p^2 dp}{(1-\alpha p +\alpha^2 p^2)^4}\\
& = \frac{L^3}{2\pi^2}\times\frac{32\pi\sqrt{3}+81}{243 \alpha^3}~.
\ee
\par\noindent
In Fig. 3., we plot the number of states, i.e., $N$, as a function of the momentum, i.e., $p$, of the single massless particle, when computed in different versions of GUP. The effect of the different weight factors on $N$ is easily seen.
\begin{figure}[h!]
  \includegraphics[width=0.5\textwidth]{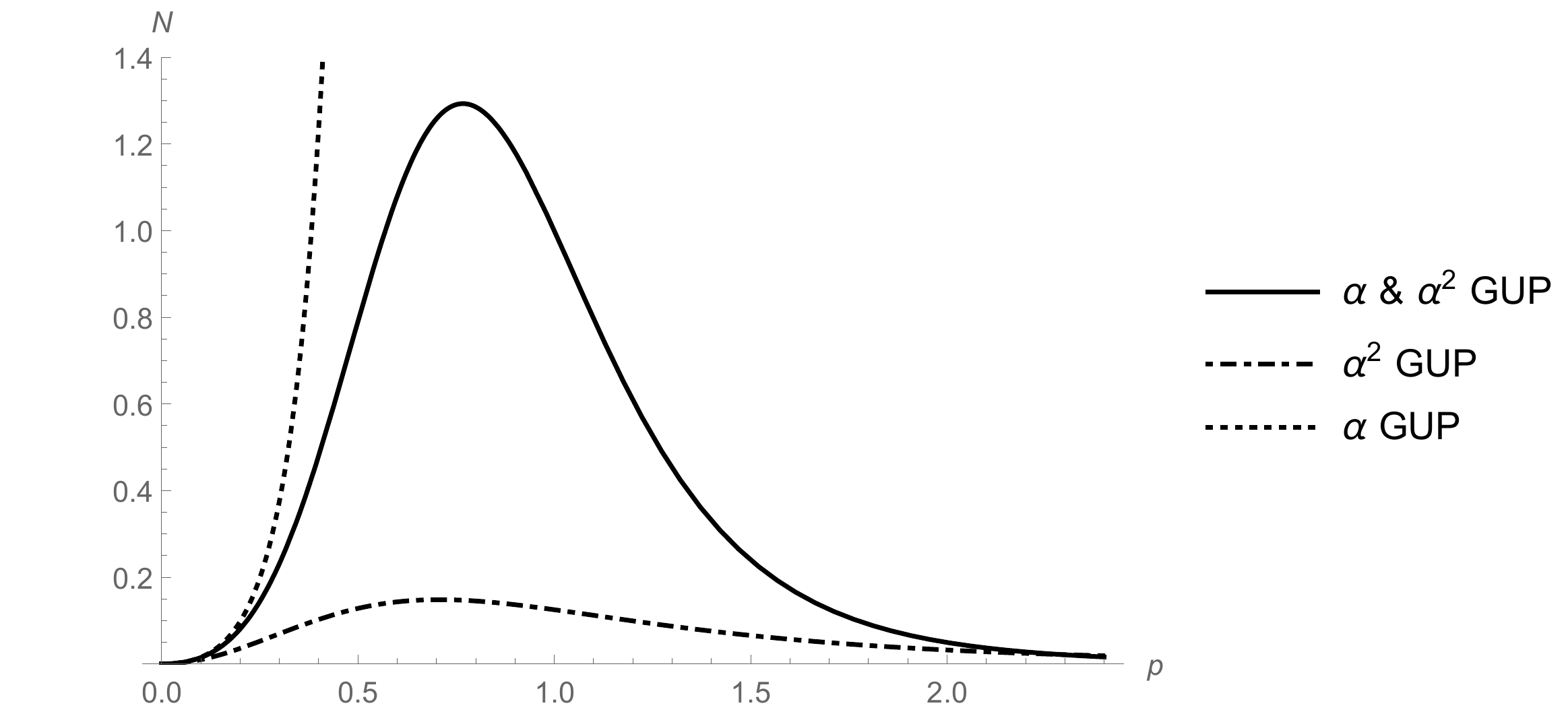}
  \caption{The effect of different weight factors on the calculation of the number of states upon considering LGUP, QGUP and LQGUP in D=3. We have set $\alpha=1$.}
  \label{fig.3}
\end{figure}
It is also noteworthy that   the number of states of the LQGUP is higher than the suppressed one of the QGUP [21], due to the contribution of the linear term, i.e., $\alpha$ term. However,  the number of states of LQGUP remains convergent compared to that of the LGUP.
\par\noindent
Now, we compute the corresponding energy of the black body massless radiation
\be\label{Energy}
E = & \frac{L^3}{2\pi^2}\int\limits_0^{\infty} \frac{p^3 dp}{(1-\alpha p +\alpha^2 p^2)^4}\\
& = \frac{L^3}{2\pi^2}\times\frac{28\pi\sqrt{3}+81}{243 \alpha^4}~.
\ee
It should be pointed out that the energy of the black body massless radiation has similar behavior with  $\Lambda(m=0)$ with respect to the GUP parameter $\alpha$. This is easily seen since both Eq.(\ref{Lambda0})  and Eq.(\ref{Energy}) are inversely proportional to $\alpha^4$. Thus, it is expected that the plot of energy of the black body massless radiation given by Eq.(\ref{Energy}) as a function of a function of the momentum, i.e., $p$, of the single massless particle will be very similar to Fig. 2.
\par\noindent
At this point, it is very important to introduce the following functions
%
%
%
 %
 %
 %
\be \label{spectral}
& g_0(w,T) \equiv \frac{(w/w_{\alpha})^3}{e^{(w/w_{\alpha})(T_{\alpha}/T)} - 1 }\\
& g_{\alpha}(w,T) \equiv \frac{ 1 }{ [1-w/w_{\alpha}+(w/w_{\alpha})^2]^4 }\,g_0(w,T)
\ee
with $w$ to be the frequency of the spectral function, and the constants $\displaystyle{w_{\alpha} \sim \frac{1}{\alpha}}$, and $\displaystyle{T_\alpha \sim \frac{1}{k_{B}\alpha}}$. These functions will help to compute the energy of the black body massless radiation in a curved spacetime when the LQGUP is taken into consideration.
%
%
%
%
%
\section{LQGUP and Massless Particles in Curved Spacetime}
%
%
%
%
%
\par\noindent
In this section, we expand the analysis of Ref. \cite{Li:2009bh}. In particular, in Ref.  \cite{Li:2009bh} the total energy density of massless particles was computed  in the context of QGUP and using the unit volume of phase space obtained in Ref.  \cite{Chang:2001bm}.  Now, we employ Eq.(\ref{volume}) in such a way  that at the WKB level, the norm of 3-momentum vector of a massless particle reads
\be
p^2=p_i p^i = \frac{w^2}{f(r)}
\ee
where $f(r)\equiv -g_{tt}$ is the metric element of any static spherically symmetric metric like the Schwarzschild, Reissner-Nordstr\"{o}m, Bardeen, Hayward, and (anti-)de Sitter spacetime background, or any combination of them. If we set $D=3$, then the total energy density for all frequencies will be
\be\label{rho}
\rho(f,\beta)=\gamma \int\limits^{\infty}_0 \frac{f^2 ~ w^3}{2\pi^2 (f-\alpha \sqrt{f} w + \alpha^2 w^2)^4} \times \frac{1}{e^{\beta w}\pm 1}dw
\ee
where  $f=f(r)$,  $\gamma$ is the spin degeneracy, the negative sign in the denominator stands for the massless bosons, while the positive sign stands for the massless fermions.
Upon considering the change of variables $x=\beta w/2\pi$ and $T(r)=1/(\beta \sqrt{f})$, with  $T(r)$ to be the local temperature in a curved spacetime and $\beta$ is the reciprocal temperature \footnote{Henceforth, the $\beta$ will be the reciprocal temperature, and not  the GUP parameter $\beta$ that appears in Ref. \cite{Chang:2001bm}.}, Eq.(\ref{rho}) becomes
\be\label{rhoT}
\rho(x,T)=8\pi^2\gamma T^4 \int \limits^{\infty}_0 \frac{x^3}{(1-ax+a^2 x^2)^4} \times \frac{1}{e^{2\pi x}\pm 1}dx
\ee
\par\noindent
with $a=2\pi\alpha T$. This integral is not easy to be exactly solved, but it is indeed a convergent integral. So upon expanding the denominator up to $\mathcal{O}(a^3(\alpha))$ and setting $x=s/2\pi$, we get
\be\label{BosonsFermions}
\rho(s,T) \sim 4\pi\gamma T^4 \int \limits^{\infty}_0 \Big[\frac{1}{(2\pi)^3}\frac{s^3}{e^{s}\pm 1} + \frac{4a}{(2\pi)^4}\frac{s^4}{e^{s}\pm 1} + \frac{6a^2}{(2\pi)^5}\frac{s^5}{e^{s}\pm 1} \Big] ds~.
\ee
For the case of massless bosons, we use the Riemann zeta function
\be
\zeta (s) = \frac{1}{\Gamma (s)} \int\limits_0^{\infty} \frac{x^{s-1}}{e^x-1} dx \qquad \text{where} \qquad s\in\{4,5,6\}
\ee
and, for the case of massless fermions, we use Dirichlet eta function
\be
\eta (s) = \frac{1}{\Gamma (s)} \int\limits_0^{\infty} \frac{x^{s-1}}{e^x+1} dx \qquad \text{where} \qquad s\in\{4,5,6\}~.
\ee
Finally, we provide the Figs. 4, 5, 6, and 7 to demonstrate and compare the effect of HUP, QGUP, and  LQGUP on the total energy density of massless particles. For fixed $\alpha$ and $T(r)$, we assume $a$ to be small compared with $x$. When $\alpha$ and $T(r)$ conspire to render a very diminutive values of $a$ for general static spherically symmetric spacetime, as in Fig. 6 and Fig. 7, we notice that GUP correction tends to be HUP, as expected, for both massless bosons and fermions. Since HUP dies slower than LQGUP,  we agree with Chang {\it et al.} that the distortion to the black body radiation is undetectable, and the spectrum of the Cosmic Microwave Background (CMB) stays unaffected too.
As an example for the effect of LQGUP on the radiation distribution of massless particles, we discuss in Ref. \cite{Vagenas:2019rai} the case of an \emph{ultracold} RNdS-like spacetime and its corresponding massless \emph{charged} particles.
\begin{figure}[h!]
\centering
\begin{minipage}[t]{0.48\linewidth}
\includegraphics[width=\linewidth]{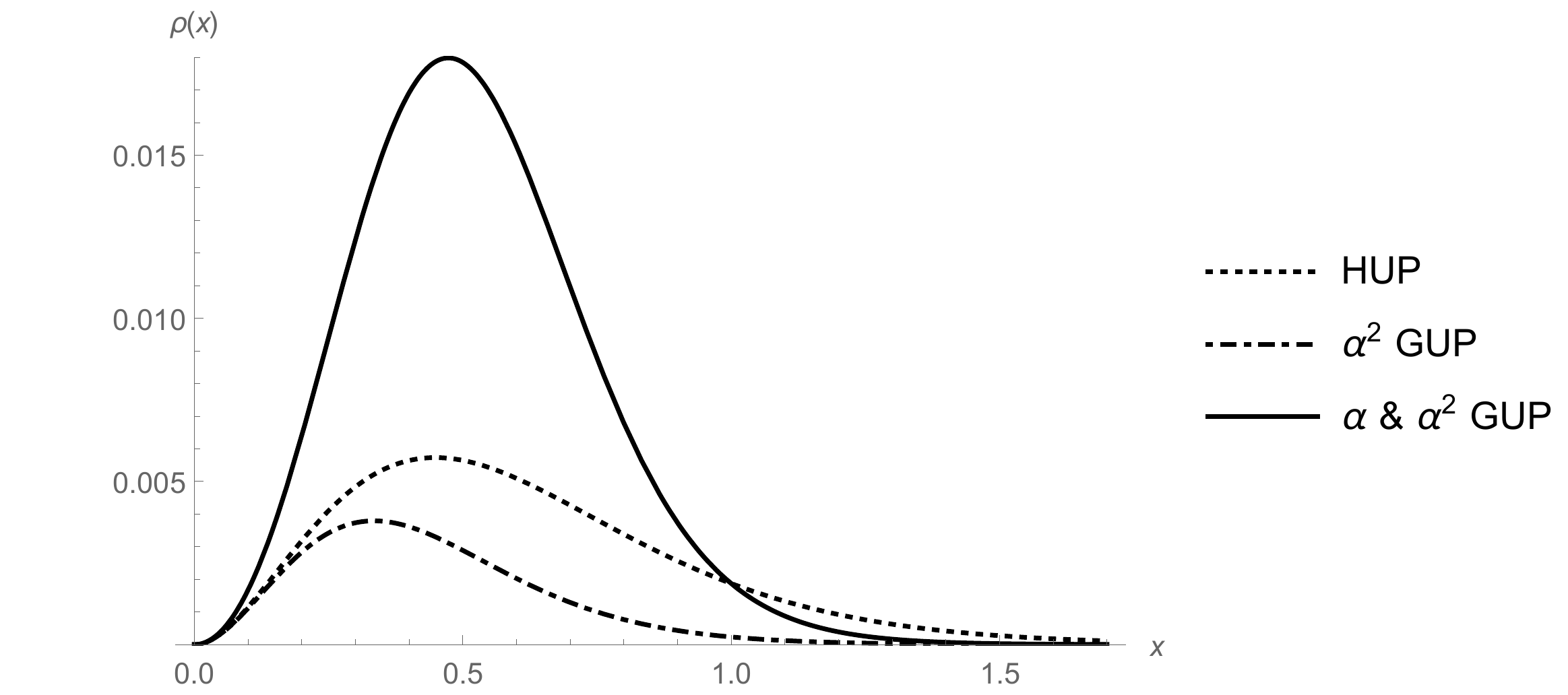}
\caption{The total energy density $\rho(x)$ versus the variable x for massless bosons in a general static spherically symmetric spacetime, where $a=1$.}
\label{fig.4}
\end{minipage}\hfill
\begin{minipage}[t]{0.48\linewidth}
\includegraphics[width=\linewidth]{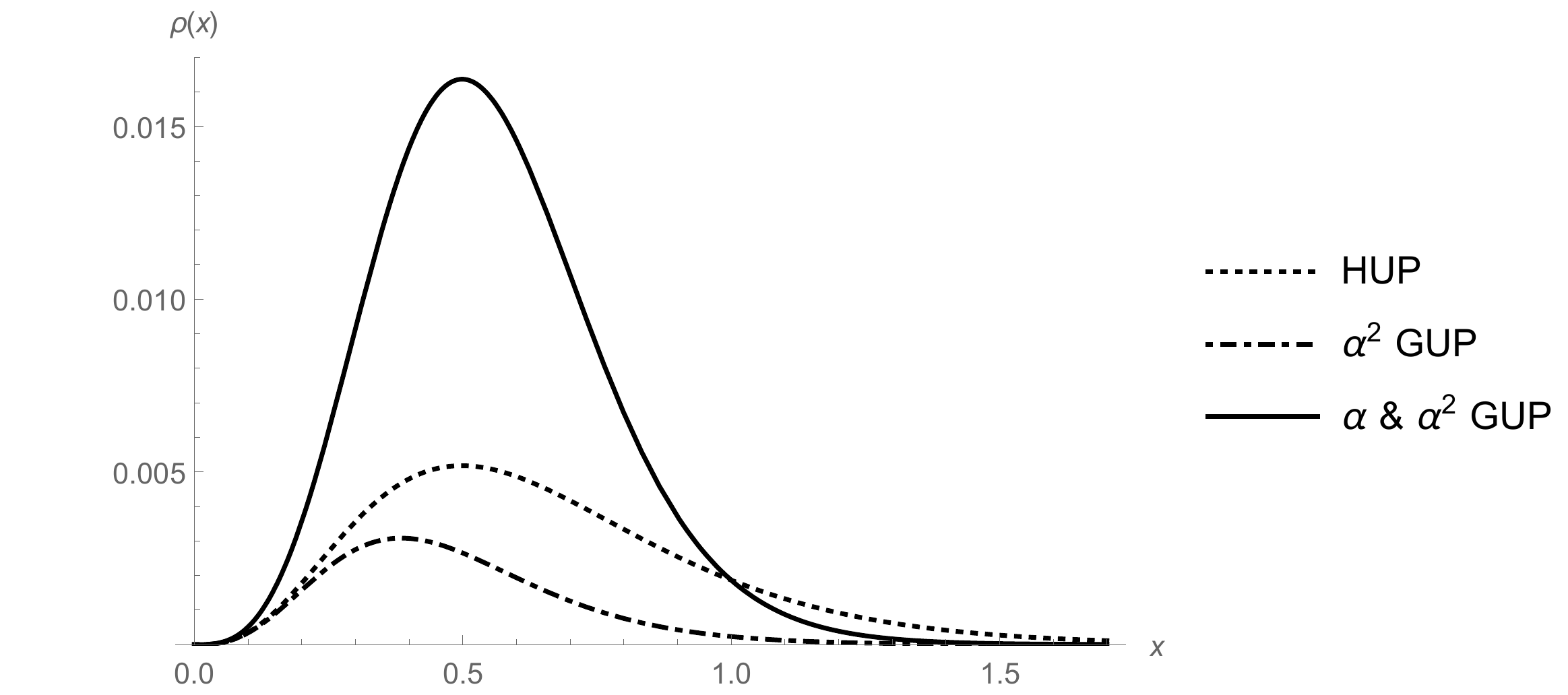}
\caption{The total energy density $\rho(x)$ versus the variable x for massless fermions in a general static spherically symmetric spacetime, where $a=1$.}
\label{fig.5}
\end{minipage}
\end{figure}
\begin{figure}[h!]
\centering
 \begin{minipage}[t]{0.48\linewidth}
	\includegraphics[width=\linewidth]{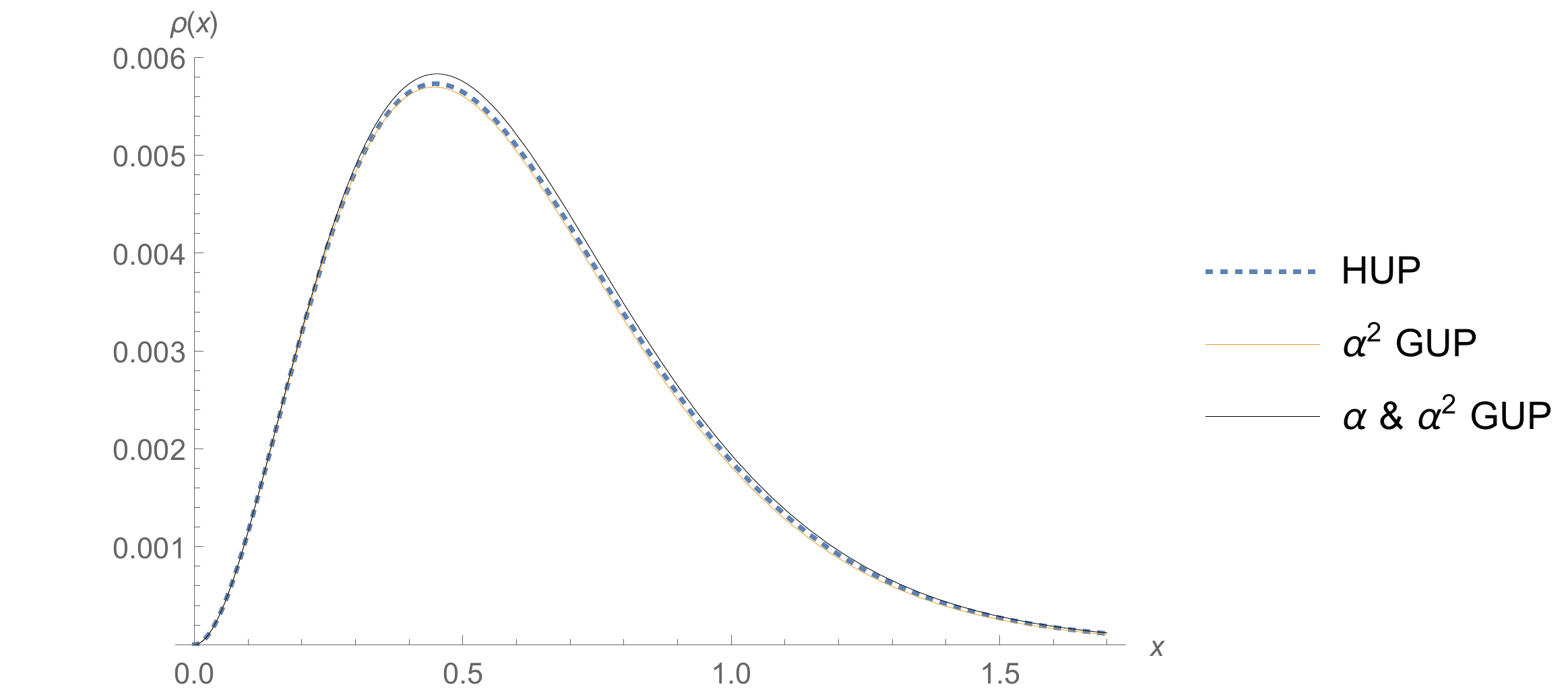}
	\caption{The total energy density $\rho(x)$ versus the variable x for massless bosons in a general static spherically symmetric spacetime, where $a=0.01$.}
	\label{fig.6}
  \end{minipage}\hfill
 \begin{minipage}[t]{0.48\linewidth}
	\includegraphics[width=\linewidth]{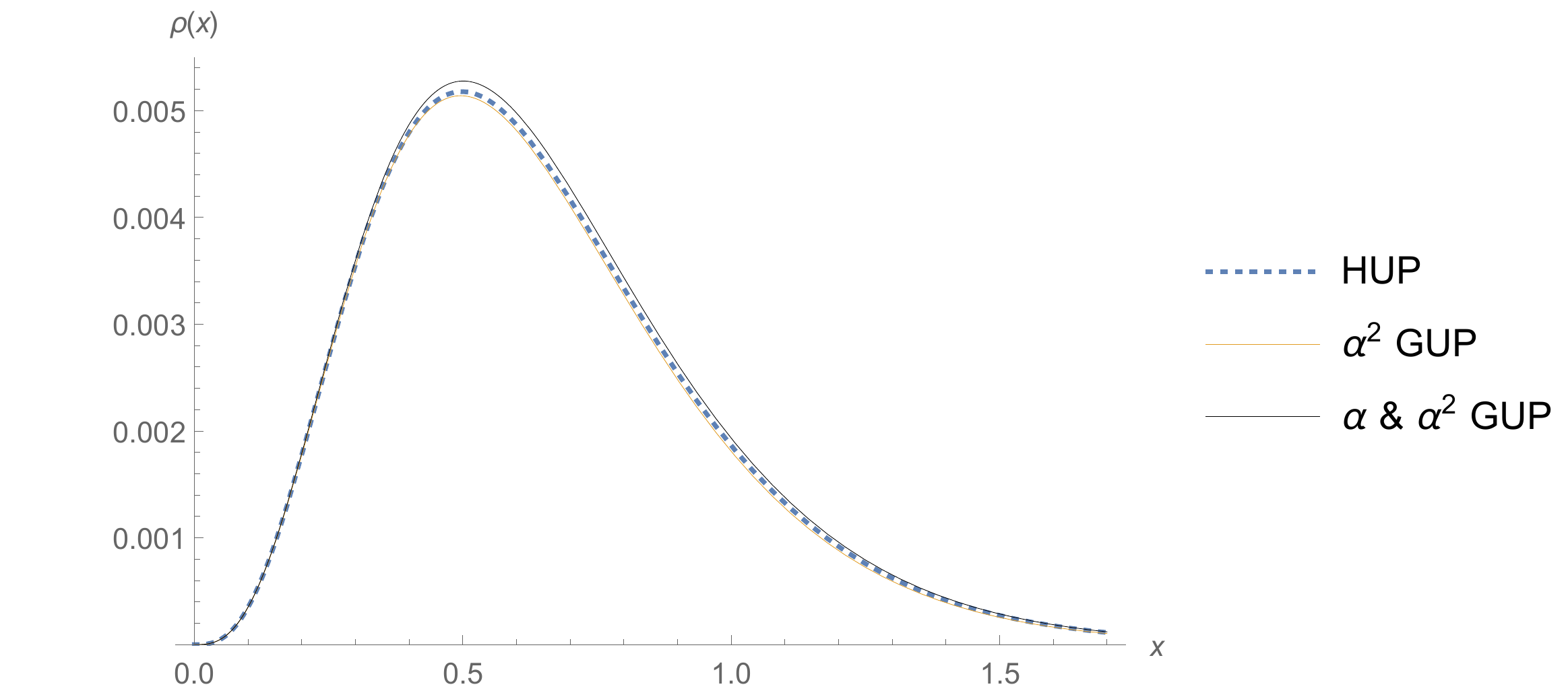}
	\caption{The total energy density $\rho(x)$ versus the variable x for massless fermions in a general static spherically symmetric spacetime, where $a=0.01$.}
	\label{fig.7}
  \end{minipage}
\end{figure}
%
%
%
%
%
%
\section{LQGUP effect on Brick Wall Entropy}
%
%
%
%
\par\noindent
Motivated by Ref. \cite{Li:1992sy}, Li  calculated,  in the context of QGUP,  the energy density of the black body radiation as follows \cite{Li:2002xb} 
\bea
u  &=& \int_{0}^{\infty} \frac{\omega^3 d\omega}{(e^{\beta \omega}-1)(1+\alpha^2
\omega^2)^3} \\
& = & \beta^{-4}\int_{0}^{\infty} \frac{x^3
dx}{(e^{x}-1)(1+ax^2)^3}
\eea
\par\noindent
where $a=\left(\alpha/\beta\right)^{2}$ and $x=\beta\omega$. The above integral was solved asymptotically first by setting the HUP condition, namely $\alpha\to 0$ which means the temperature is much less than Planck temperature, and then by setting the upper bound condition of the  energy density. Thus, we adopt the same analysis here except that we introduce our new weight factor. Similar to the result we obtained from Eq.(\ref{Lambda0}), the \emph{numerical} correction that comes from LQGUP is expected to be much larger than that of QGUP. However, it will not substantially change the convergent behavior of the function as we have seen before. The upper bound of energy density is given by 
\bea\label{bound}
u&<&\beta^{-4}\int_{0}^{\infty} \frac{x^2 dx}{(1-\frac{\alpha x}{\beta}+\frac{\alpha^2
x^2}{\beta^2})^4}\\
&=&\frac{32\pi\sqrt{3}+81}{243\alpha^3}\beta^{-1}
\eea
\par\noindent
where the inequality comes from the  fact that $(e^x-1)>x$ which means that when the temperature is higher than the Planck temperature, the state equation of the thermal radiation is different from that of HUP, i.e., $u\sim
\beta^{-4}$  \cite{Li:2002xb} . From Eq.(\ref{volume}) and Eq.(\ref{spectral}), the number of quantum states with energy less than $\omega$ is given by
\be\label{gom}
g(\omega)& = \frac{1}{(2\pi)^3}\int \frac{dr~d\theta~d\varphi~dp_r~dp_{\theta}~dp_{\varphi}}{(1-\alpha\omega/f^{1/2}+\alpha^2\omega^2/f)^4}\\
& = \frac{1}{(2\pi)^3}\int \frac{dr~d\theta~d\varphi}{(1-\alpha\omega/f^{1/2}+\alpha^2\omega^2/f)^4}\int\frac{2}{f^{1/2}}\left[\frac{\omega^2}{f}-\frac{1}{r^2}p_{\theta}^2-\frac{1}
{r^2\sin^2\theta}p_{\varphi}^2\right]^{1/2}dp_{\theta}~dp_{\varphi}\\
& = \frac{4\pi\omega^3}{3(2\pi)^3}\int \frac{r^2dr}{f^2(1-\alpha\omega/f^{1/2}+\alpha^2\omega^2/f)^4}\int \sin\theta~d\theta~d\varphi\\
& = \frac{2\omega^3}{3\pi}\int \frac{r^2dr}{f^2(1-\alpha\omega/f^{1/2}+\alpha^2\omega^2/f)^4}~
\ee
\par\noindent
and when $\alpha\to 0$, Eq.(\ref{gom}) goes back to the standard expression in the HUP limit.  Furthermore, the free energy reads
\be
F(\beta) & = \frac{1}{\beta}\int
dg(\omega)\ln(1-e^{-\beta\omega})\\
& = -\int_{0}^{\infty} \frac{g(\omega)d\omega}{e^{\beta \omega}-1}\\
& = -\frac{2}{3\pi}\int_{r_0}\frac{r^2dr}{f^2}\int_{0}^{\infty}\frac{\omega^3d\omega}
{(e^{\beta \omega}-1)(1-\alpha\omega/f^{1/2}+\alpha^2\omega^2/f)^4}~.
\ee
\par\noindent
Therefore, the entropy is written in the from
\be
S & = \beta^2\frac{\partial F}{\partial \beta}\\
& = \frac{2\beta^2}{3\pi}\int_{r_0}\frac{r^2dr}{f^2}\int_{0}^{\infty}\frac{e^{\beta\omega}
\omega^4d\omega}{(e^{\beta \omega}-1)^2(1-\alpha\omega/f^{1/2}+\alpha^2\omega^2/f)^4}\\
& = \frac{2\beta^{-3}}{3\pi}\int_{r_0}\frac{r^2dr}{f^2}\int_{0}^{\infty}\frac{x^4dx}
{(1-e^{-x})(e^{x}-1)(1-\frac{\alpha x}{\beta f^{1/2}}+\frac{\alpha^2 x^2}{\beta^2 f})^4}~.
\ee
\par\noindent
In the light of the following inequalities
\be
& 1-e^{-x}>\frac{x}{1+x}\\
& e^{x}-1>x
\ee
\par\noindent
the entropy satisfies the inequality
\be
S & < \frac{2\beta^{-3}}{3\pi}\int_{r_0}\frac{r^2dr}{f^2}\int_{0}^{\infty}\frac{(x^3+x^2)dx}
{(1-\frac{\alpha x}{\beta f^{1/2}}+\frac{\alpha^2 x^2}{\beta^2 f})^4}\\
& = \frac{2\beta^{-3}}{3\pi}\int_{r_0}\frac{r^2dr}{f^2}\left[\frac{28\pi\sqrt{3}+81}{243(\alpha/\beta)^4}f^2+\frac{32\pi\sqrt{3}+81}{243(\alpha/\beta)^3}f^{3/2}\right]\\
& = \frac{2}{3\pi}\frac{28\pi\sqrt{3}+81}{243\alpha^4}\beta\int_{r_0}r^2dr+\frac{2}{3\pi}\frac{32\pi\sqrt{3}+81}{243\alpha^3}
\int_{r_0}\frac{r^2dr}{f^{1/2}}~.
\ee
\par\noindent
Since we consider the upper bound, we only want to get contribution from the domain close to the horizon, $[r_0, r_0+\epsilon]$, that corresponds to the minimal length $\sim \alpha$, i.e., it is just the neglected vicinity in the Brick Wall model \cite{tHooft:1984kcu, Demers:1995dq}. Therefore, we have
\be
2\alpha & = \int_{r_0}^{r_0+\epsilon}\frac{dr}{\sqrt{f}}\\
& \sim  \int_{r_0}^{r_0+\epsilon}\frac{dr}{\sqrt{2\kappa(r-r_0)}}\\
& \sim \sqrt{\frac{2\epsilon}{\kappa}}
\ee
where $\kappa=2\pi\beta^{-1}$ is the surface gravity at the horizon of black hole.
Finally, the entropy is written as
\be
S & \sim
\frac{2}{3\pi}\frac{(28\pi\sqrt{3}+81)}{243\alpha^4}\beta r_{0}^2\epsilon+\frac{2}{3\pi}\frac{(32\pi\sqrt{3}+81)}{243\alpha^3}
2r_0^2\alpha\\
& \sim 0.239\frac{A}{\alpha^2}~.
\ee
\par\noindent
It is evident that the entropy, $S$, is proportional to the black hole horizon area $A=4\pi r_{0}^2$ and, in addition, the entropy is less than $A/4\alpha^{2}$, as expected.  Furthermore, it is also anticipated from the previous sections that by  introducing the linear term to the QGUP it will cause the convergent QGUP energy distribution, and, consequently, the entropy to significantly increase \footnote{In Ref. \cite{Li:2002xb}, $\lambda = \alpha^2$.}. However, the convergent behavior remains the same. So, in contrary to LGUP effect on entropy \cite{Ali:2011ap}, we agree with QGUP of Ref. \cite{Li:2002xb} that LQGUP does not need any cutoff near the horizon. The last result emphasizes that minimal length contributes to black holes such that it may provide simpler interpretation without introducing a \emph{divergent} assumption like the Brick Wall model \footnote{The claim that there is no need for the introduction of the Brick Wall model in order to keep under control the divergences appearing when one approaches the horizon, was also supported  in the framework of Gravity's Rainbow \cite{Garattini:2009nq, Garattini:2017bgx}.}.
%
%
%
%
%
\section{Conclusion}
%
%
%
%
%
\par\noindent
Motivated by the unexpected ramification upon employing LQGUP to no-cloning theorem \cite{Vagenas:2018pez}, we discuss the consequences of applying the  LQGUP on the characteristics of the momenta distribution in phase space, particularly IR/UV behaviors. It is shown that in QGUP of Ref. \cite{Chang:2001bm} that  the UV behavior is convergent, while it is divergent in LGUP of Ref. \cite{Ali:2011ap}. So we reconcile them through LQGUP. Upon employing LQGUP on equations of motion, we agree with  Ref. \cite{Ali:2011ap} that the acceleration is no longer mass-independent, and hence, the equivalence principle is dynamically violated. Then, we modify the Liouville theorem in the presence of LQGUP and show that the weight factor has power $(D+1)$ as in Ref. \cite{Ali:2011ap} and a quadratic term as in Ref. \cite{Chang:2001bm}, but with a numerical factor that depends on $D$. Next, we encounter the cosmological constant problem. We deduce that LQGUP has similar convergent form of that in Ref. \cite{Chang:2001bm} rather than the divergent behavior of that in Ref. \cite{Ali:2011ap}. However, it still can not resolve the cosmological constant problem as in Ref. \cite{Chang:2001bm} due to the $\Lambda(0)\sim 1/\alpha^4$ together with the fact that $\alpha^2\sim M_p$. After that, we compare the different consequences of each corresponding weight factor of LGUP, QGUP, and LQGUP on the number of massless bosonic states of black body radiation. The LQGUP shows a convergent behavior similar to that of QGUP despite it is much larger in the number of states. That larger number is due to the linear term, which by its  own has divergent behavior as in Ref. \cite{Ali:2011ap}. It is obvious that the linear and quadratic terms together conspire to give such behavior. Moreover, we show how that reflects on the calculation of the energy distribution and gives the same behavior. Later, we introduce the gravitational effects on the energy of  massless bosons and fermions. We notice the agreement with Ref.  \cite{Chang:2001bm} on the unaffected CMB and undistorted radiation of black body, and that is guaranteed by the faster decay of LQGUP compared with HUP. Furthermore, we get the bosons' behavior to be with slightly higher energy density than that of fermions, as expected. HUP, QGUP, and LQGUP get very close to each other for very small values of $\alpha$, as expected too. Finally through LQGUP and QGUP of Ref. \cite{Li:2002xb} but not the LGUP of Ref. \cite{Ali:2011ap}, we agree that minimal length would ``guard'' the entropy of black holes so that there is no need for any Brick Wall  model. 


\end{arabicfootnotes}
\end{document}